\documentstyle[multicol,aps,epsfig]{revtex}

\begin{document}
\draft
\title{Microgels and fractal structures at interfaces and surfaces}
\author{Thomas A. Vilgis and Michael Stapper}
\address{Max-Planck-Institut f\"ur Polymerforschung,
Postfach 3148, 55021 Mainz, Germany}
\date{\today}
\maketitle

\begin{abstract}
The behavior of microgels near surfaces and their adsorption
is studied by simple scaling theory. Two different types of microgels can be
studied, i.e., fractal type microgels and randomly crosslinked polymer
chains. In the first case the gel can be described mainly by introducing a
spectral dimension. The second type requires more attention and uses the
number of crosslinks as parameter. The main result is that soft gels with
weakly coupled crosslinks and a low number of 
crosslinks adsorb much better than
hard gels, with many crosslinks. Similar results for fractal gels and branched
polymer are presented. Fractal gels with low connectivity adsorb easier than
gels with a large connectivity dimension. We discuss also
consequences on surface
protection by microgels.

\end{abstract}

\section{Introduction}

The theory of polymers near surfaces is a very important subject for
theoretical investigation. The main reason is the very broad and deep
theoretical research possible for  these types of systems. 
Modern theories have been
developed  and brought to a very high standard \cite{eisenriegler}. 
On the other
hand there is a large demand on 
practical interest on studying polymers near surfaces. One of them is surface
protection. The theory on polymer brushes \cite{brush}
is a typical example of these type
of applications. Polymers are attached to surfaces to protect the surface from
further adsorption of, e.g., biological active molecules such as
proteins. These molecules need to be pretended to from all kind of
surface interactions in
order to save their biological function \cite{mpiruhe}. The (technical)
problem with polymer brushes is, that a large surface coverage is needed to
protect the surface in a most effective way. Polymers attached chemically or
physically at one end at the surface form brushes, i.e., the chains are
extended and the  brush height $h$ follows a scaling law 
$h \simeq \sigma^{-1/3} N$ (in good solvent), 
where $\sigma$ is the area per chain an is thus
related to the grafting density. $N$ the
degree of polymerization of the chains. The problem is to set up a small value
of $\sigma$, or, correspondingly 
a large grafting density. For low values of the grafting
density the chains behave as "mushrooms" and the surface protection is
incomplete. 

In earlier papers it has been shown by one of us, that the use of branched
chains in much more effective \cite{viharo,vilharoben,vizhul}. 
Chains, polymers and
polymeric fractals with a larger connectivity seem to be more appropriate to
protect surfaces more effectively. Indeed due to their connectivity their
occupied area is larger and it turns out that these systems behave more like
single chain fractals. Typical many body effects, such as occur in conventional
polymer brushes do not play a significant role.

In the present paper we suggest a different route of surface
protection by using microgels and branched structure. 
Microgels had become an important tool in designing polymeric nano
structures. These systems can be synthesized \cite{a1,a2,a3,a4} with different
structures. Indeed the
structure of these microgels can range from a fractal state, i.e., a branched
self-similar 
polymer with a large connectivity, up to almost hard and highly crosslinked
spheres. Such systems are well designed to study the transition form polymer
to colloid behavior by variation of the structure and the crosslinking state.

We have shown earlier \cite{vilgis,viharojcp}, 
that fractal polymers and gels can interpenetrates each
other and screen excluded volume forces, wherever their connectivity is
low. In terms of the spectral dimension this is the case, when it is lower than
a critical value, i.e. $D_{\rm c}= 6/5$. Thus fractals with lower spectral
dimensions screen their excluded volume, whereas polymeric fractals with
larger spectral dimension saturate. Then they form  soft balls, which cannot
interpenetrate each other and are
well separated from each other \cite{vilgis,viharojcp}. 
This state has an analogon in the case of linear polymers. Polymer melts in
two dimensions correspond to a saturated state. The individual chains are
separated from each other and form  on average disks on a hexagonal lattice
\cite{kremer}. 

For surface protection in three dimensions it would thus be more effective to
use fractal microgels with a large connectivity. Then the surface coverage is
ruled by the single gel behavior. Alternatively crosslinked gels can be used to
have the same effect. Sufficiently crosslinked gels cannot interpenetrate each
other. The adsorption behavior can be studied then by the single gel
adsorption. The significant parameters are either the connectivity (spectral
dimension $D$) or the crosslink number $M$. In the following we emphasize
mainly on the crosslinked gels rather on the self similar connected polymeric
fractals. Nevertheless we will  consider both cases below.

The paper is organized as follows. In section \ref{2} we introduce the model
of self crosslinked polymer chains, which form the microgels. In section
\ref{3} we repeat briefly the scaling behavior of ideal microgels before 
we consider  in section \ref{4} the effects of excluded volume in the bulk.
Sections \ref{5} and \ref{6} treat the adsorption behavior of the gels close
to attracting walls using simple scaling arguments. In section \ref{5} we will
also make some remarks on fractal type gels.

\section{The model}\label{2}
Flexible interacting macromolecule modeled usually  by the
Edwards Hamiltonian in three dimensional space. The Edwards Hamiltonian
consists of two parts, i.e.,the Gaussian connectivity of monomers
\begin{equation}
\nonumber
H_{\rm W}=\frac{3}{a^{2}}\sum_{i=1}^{N}({\bf R}_{i}-{\bf R}_{i-1})^{2}
\end{equation}
and the self avoidance between monomers
\begin{equation}
\nonumber
H_{\rm I}=v\sum_{0\le i\le j}^{N}\delta ({\bf R}_{i}-{\bf R}_{j})
\end{equation}
Thereby  $v>0$ is the  excluded volume of the monomers, $N$ the 
degree of polymerization, and $a$ the Kuhn length. The chain configurations 
are
determined by   monomer coordinates ${\bf R}_{i}$, where $i$ labels all
monomers $1 \leq i \leq N$.
The Edwards Hamiltonian is sufficient to describe a free self avoiding walk
chain. To study the properties of a microgel crosslinks have to be introduced.
The most obvious statistical representation of a microgel is a self
crosslinked single chain. Such a situation has been studied many years ago by
the Manchester group in three papers \cite{sfemanch}. The
static  properties have been computed
by variational techniques. In this paper 
we choose  a different route. Let us therefore 
consider a  microgel as a self  crosslinked polymer of roughly
spherical structure, which can be visualized as given in Fig.(\ref{one}).

\begin{center}
\begin{minipage}{10cm}
\begin{figure}
\epsfig{file=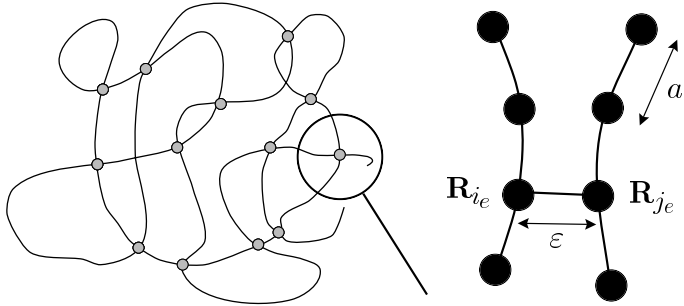,width=9cm}
\vspace{0.5cm}
\caption{\label{one}\small{A self crosslinked chain forming a microgel. The crosslinks are
  represented by distance constraints. Two segments are held a distance 
$\epsilon$ apart. The hard microgel is given by $\epsilon \to 0.$}}
\end{figure}
\end{minipage}
\end{center}

In this paper we want to describe the crosslinks such that a continuous
transition between the free chain and the fully crosslinked state can be
represented in the same model. This has been motivated by our earlier work,
where soft crosslinks have been introduced. The method of using
soft crosslinks is able to interpolate from the free chain to the hard
microgel. The state in between is a new kind of branched chain, whose scaling
properties have been already described \cite{solf1,solf2,solf3,tus}. 
To be more
precise let us introduce $M$ permanently crosslinked monomers, where
 each of them is characterized by pair of randomly  chosen 
monomer coordinates $i_{e},j_{e}$ that form a permanent crosslink.
In fact, the whole set of crosslinks ${\sf C}=\{i_{e},j_{e}\}_{e=1}^{M}$ 
determine the  random connectivity of the micro network. These definitions
and proposals allow us to formulate the partition function of the microgel,
i.e., 
\begin{equation}
\nonumber
Z({\sf C})=\int_{V} \prod_{i=0}^{N}d{\bf R}_{i}{e}^{-(H_{W}+H_{I})}
\prod_{e=1}^{M}\delta ({\bf R}_{i_{e}}-{\bf R}_{j_{e}})
\end{equation}
The partition function describes the Gaussian network with self avoiding
interactions, and takes into account the total connectivity of the certain
crosslink configuration {\sf C}. If the delta constraint for the permanent
crosslinks
is represented by a
soft Gaussian function, i.e., 
$\delta (x) = 
\lim_{\varepsilon \to 0}({\rm const}/\epsilon) \exp(-(3/\epsilon^{2})x^{2})$, the
problem can be solved exactly. Thus we model the
$\delta$-function by Gaussian distribution with width $\epsilon $
in limit $\epsilon \to 0$ and formulate the Hamiltonian of the crosslinked
chain by
\begin{eqnarray}
\nonumber
H_{W}&=&\frac{3}{a^{2}}\sum_{i=1}^{N}({\bf R}_{i}-{\bf R}_{i-1})^{2}
      +\frac{3}{\epsilon^{2}}\sum_{e=1}^{N}
        ({\bf R}_{i_{e}}-{\bf R}_{j_{e}})^{2}\\
&&      + v\sum_{0\le i\le j}^{M}\delta ({\bf R}_{i}-{\bf R}_{j})
\end{eqnarray}
\section{Microgels without interactions}\label{3}
Although we have shown in previous publications that the partition function
can be solved for any value of $\epsilon$ exactly it is useful to rederive the
results previously by the use of scaling arguments. These consideration will
reproduce the exact results apart from prefactors. To do so it is useful to
recall that the classical random walk contains two elastic contributions, one
for stretching and one for compression. The addition of both
yield a free energy $F \sim \frac{R^{2}}{a^{2}N}+\frac{a^{2}N}{R^{2}}$
and by minimization the size of Gaussian chain $R\simeq a N^{1/2}$ is
recovered. Let us shortly repeat the results for later use. For the case of
soft crosslinks, i.e., whenever $\epsilon$ is within the range 
$ a^{2} << \epsilon^{2} <<  a^{2}N M$
we have for the relevant part of Flory free energy 
\begin{equation}
F\simeq \frac{N a^{2}}{R^{2}}+\frac{1}{\epsilon^{2}}M R^{2}
\end{equation}
and minimization yields
the branched polymer regime
\begin{equation}
R\simeq a\left( \frac{\epsilon }{a}\right) ^{1/2}\left( \frac{N}{M}%
\right) ^{1/4}
\label{soft}
\end{equation}
The appearance of the typical branched polymer exponent $\nu = 1/4$ is not
surprising in the ideal case, since the constraint can be visualized as
springs. Then the connectivity is changed. The branched polymer regime in the
range of $a << \epsilon << a\sqrt{NM}$ was confirmed also by the exact
solution of the crosslinked chain problem, see \cite{solf2}. Moreover it
agrees also with the corresponding branched polymer \cite{tus}

In the case of hard crosslinks, whenever $\epsilon \simeq 0\; {\rm or}\; a$ 
the crosslink term must be differently estimated. 
To do so the crosslink term can be estimated by the size of the random
walk through crosslinks. Then the relevant part of Flory free energy is then
given by
\begin{equation}
F\simeq \frac{a^{2}N}{R^{2}}+M\frac{R^{2}}{a^{2}\left( 
\frac{N}{M}\right) }
\end{equation}
Minimization of the free energy provides the size of the microgel by
\begin{equation}
R\simeq a\left( \frac{N}{M}\right) ^{1/2}
\end{equation}
as also given by the exact results \cite{solf2}.  Formally the latter result
can be found from a special choice for $\epsilon$ from the corresponding
result for soft gels (\ref{soft}), but note that the way of estimating the free
energy contribution of the crosslinks is estimated very differently. Thus there
can be a different prefactor, which is not accessible by scaling. The exact
values for the radius of gyration of the non-interacting but crosslinked
chains have been computed exactly in \cite{solf2}, where the numerical
prefactors can be found.
\section{Microgels with excluded volume }\label{4}
Although we have been able to compute the size of the microgel exactly
whenever the interactions are not present the self avoiding case appears very
difficult.
The only possibility for the problem at this stage is to use 
Flory estimates for size $R$. Let us first consider the case of soft 
crosslinks. The distance constraint that forces two arbitrarily chosen polymer
segments together shrinks the chain. The shrinkage costs entropy penalty which
balances with the distance constraint. The use of the pseudo potentials allows
us, however, to cast this in a simple Flory free energy  
to
\begin{equation}
F\simeq M\frac{R^{2}}{\epsilon ^{2}}+a^{3}\frac{N^{2}}{R^{3}}
\end{equation}
and minimization yields the size of the swollen soft microgel
\begin{equation}
R\simeq a\left( \frac{\epsilon }{a}\right)^{2/5}
\left(\frac{N^{2}}{M}\right)^{1/5}
\end{equation}
The result is very intriguing. Although the Gaussian chain size scales the
same way as the branched polymer, the swelling behavior produces another
excluded volume exponent, resulting in a different swelling behavior 
as branched chains. In the latter case the swollen branched chain is
characterized by $R \propto N^{1/2}$.
On the other hand it can also be seen that the pure  scaling in terms of the
variable $N/M$ always present in the Gaussian case is destroyed. This becomes
clear, since the excluded volume introduces interactions.

A similar estimate of the size can be carried out  
in the case of hard crosslinks. Here the elastic term stemming from the
pseudo crosslink potential can be estimated the same way as in the case of non
interacting networks, which was given by a random walk through the
crosslinks. The total excluded volume energy remain the same, because it
depends only on the total amount of monomers and not on the special
connectivity. Then minimization of the free energy
\begin{equation}
F\simeq M\frac{R^{2}}{a^{2}\left( \frac{N}{M}\right) }+a^{3}\frac{N^{2}}{R^{3}}
\end{equation}
yields swollen c*-micro gels of size
\begin{equation}
R\simeq a M^{1/5}\,\left( \frac{N}{M}\right)^{3/5}
\end{equation}
The size might appear small compared to what is expected intuitively. This
comes from the fact that the crosslinks has been chosen totally randomly. In
many theories of macroscopic networks the choice of the crosslink pairs is
guided by the conformation of the excluded - or random walk chain, i.e., the
often terminated "zeroth replica" \cite{deam}.
For completeness we mention that the number of crosslinks in excluded volume
gels cannot be arbitrarily large. A natural  limit of the crosslink number is
given by the condition that the size must be larger than a fully collapsed
ball of space filling density, i.e. $R\geq aN^{1/3}$. The latter condition
yields the upper limit for the number of crosslinks to be $M\leq N^{2/3}$.

\section{{Adsorption behavior of self similar  microgels, general remarks}}
\label{5}

Let us first study the  adsorption of ideal microgels 
near a flat surface by naive scaling
arguments. To do so, we repeat the scaling idea of de Gennes for arbitrarily
flexible objects of arbitrary connectivity, but selfsimilarly linked. The 
connectivity of of such Gaussian fractal networks can be described by the
spectral dimension $D$. The total number of monomers is therefore 
$N=m^{D}$, and
their ideal size is given by $R_{0} = am^{(2-D)/2D}=aN^{1/d_{\rm f}}$, 
yielding a fractal dimension of 
$d_{\rm f} = 2D/(2-D)$. Thus $m$ counts the number of monomers through a
linear dimension through the fractal object.  
This way of description includes the wellknown cases
of linear chains for $D=1$ and for randomly branched polymers, i.e., $D=4/3$.

A simple way of looking at the adsorption conditions is to compare the free
energy penalty
of confinement of the Gaussian structure near the wall with the gain of energy
by adsorbing a certain number of monomers.
\begin{equation}
F\simeq F_{\rm conf} -n_{a}w
\end{equation}
For Gaussian chains the confinement
is simply given by by $F_{\rm conf} \sim (R_{0}/H)^{2}$.
Here we have  used the symbol $H$ for the height of adsorbed layer,
$n_{\rm a}$ is the number of adsorbed monomers, and $w$ is the 
gain of energy per $k_{B}T$. Following de Gennes book \cite{degennes} we can
reproduce the result given there for the linear ideal chain.

To do so we must first
estimate the number of adsorbed monomers at the surface, we assume that the
surface is penetrable for a moment. This yields immediately 
\begin{equation}
\label{mono}
n_{\rm a}\simeq aR^{2}\cdot \frac{m^{D}}{R^{2}H}\sim N^{}
\left( \frac{a}{H}\right) 
\end{equation}
The main problem is to estimate the confinement free energy for Gaussian
fractals with a larger connectivity compared to the linear chain. This is
definitely not just the inverse of the Gaussian free energy of stretching,
i.e., $F_{\rm stretch} = R^{2}/am^{(2-D)}$, because upon stretching only the
monomers in the shortest path are 
taking part on the deformation, whereas upon confinement the
total number of monomers are concerned. The corresponding confinement free
energy must then be of the form 
\begin{equation}
F_{\rm conf} \cong \frac{N^{}}{H^{d_{\rm f}}}
\end{equation}
This result contains the special case of linear chains, $D=1, \;d_{\rm f} =2$, 
and the latter agrees with the
classical confinement free energy.

For the ideal Gaussian structures  the height of the adsorbed
layer scales as
\begin{equation}
H\simeq a \left(\frac{1}{w} \right)^{1/(d_{\rm f}-1)}
\label{hhh}
\end{equation}
The same result can be found by a blob argument.
For selfsimilarly branched
polymers a blob model can be used \cite{vilharoben}. The manifold is confined
to a height $H$. Inside blobs of diameter $H$ the branched chain is Gaussian,
i.e., $H \simeq g^{(2-D)/2D} a$, where $g$ is the number of monomers inside
the blob. Thus the number of blobs is given by $n_{\rm b} \simeq m^{D}/g =
N^{}/H^{2D/(2-D)}$. The confinement free energy is proportional to the number
of blobs, i.e., $F_{\rm conf} \sim n_{\rm b}$. Employing the same scaling
argument as above yields immediately
$$
H \propto a (1/w)^{(2-D)/(3D-2)}
$$
which agrees with the above result, eq.(\ref{hhh}).

In any case, the above arguments only yield the
behavior of ideal chains near the surface. For excluded volume chains and
excluded volume manifolds the Gaussian elastic entropy penalty must be replaced
by the confinement energy. To do so, the manifold can be put between two plates
of distance $H$. This procedure yields similar results
along to those derived in 
de Gennes book \cite{degennes}. 

To do so, the extension of the manifold between two
parallel plates must be computed. It is given by 
$R_{\parallel} \simeq H (a/H)^{5/4} N^{(2+D)/4D}$ \cite{vilharoben}. 
This result is consistent
with the linear SAW chain between two plates. For $D=1$ the two dimensional
chain of blobs is recovered. The confinement free energy can be only a
function of the ration of the 
size of the manifold and the distance between the plates,
i.e., $(R/H)$, where $R$ is the size of the self avoiding manifold 
$R \simeq a N^{(2+D)/5D}$. The confinement free energy is then easily found 
from the condition that the free energy must be an extensive quantity. 
Thus it must scale as $F_{\rm conf} \propto N$, which yields
\begin{equation}
F_{\rm conf} \simeq kT \left( \frac{a}{H}  \right)^{5D/(2+D)}N
\end{equation}
Replacing the elastic free energy in the scaling argument above by the
confinement energy, yields the physically sensible result
\begin{equation}
H \propto a w^{- \frac{2+D}{4D-2}} 
\end{equation} 
which is now independent of the molecular weight for any manifold. Moreover
for $D=1$ the linear chain result is recovered.  Moreover the result bears an
interesting point in it: The parameter $w$ is the gain of energy at adsorption
of one segment per thermal energy $kT$. Thus it is physically reasonable that
this parameter is sufficiently small, i.e., $w<1$. Thus a significant change
of the behavior can be expected if the exponent in $H$ is larger and smaller
than one. It is interesting to note that this happens at $D=4/3$ which is
close to the spectral dimension of randomly branched chains or accidentally
for percolation clusters. Thus randomly connected manifolds of large
connectivity adsorb weaker than objects of low connectivity, as linear chains.
This is physically intuitively clear since the number of accessible sites
become smaller for increasing connectivity.  In the following section we will
use the same strategy to discuss the adsorption behavior of microgels.

\section {Confinement energy for microgels with excluded volume}\label{6}

As we have seen in the last paragraph on the previous section we have to
construct the free energy of confinement.  We use the same 
concept of gels between two plates for energy cost of squeezing. We have seen
in the first two sections that we can distinguish between soft and hard
microgels by the value of the parameter $\epsilon$. For both cases we expect
physically different behavior.

Let us first consider soft gels between plates. To do so we have to determine
the size of the gel parallel to the plates. The relevant parts of the free
energy is given by
\begin{equation}
F\simeq M\frac{R_{\parallel}^{2}}{\epsilon ^{2}}+a^{3}
\frac{N^{2}}{HR_{\parallel}^{2}}
\end{equation}
where we have argued that the elastic confinement and the "anisotropic"
excluded volume term balance each other. This yields immediately
\begin{equation}
R_{\parallel }\simeq H\left( \frac{\epsilon^{2}a^{3}}{H^{5}}\right)
^{1/4}\left( \frac{N^{2}}{M}\right)  ^ {1/4}
\end{equation}
and the corresponding confinement energy to
\begin{equation}
F_{\rm conf}\simeq \frac{N}{\sqrt{M}} \frac{\epsilon }{a}
\left( \frac{a}{H}\right)^{5/2}
\end{equation}
Note that the confinement free energy is determined by the fact that it must
be proportional to the total number of monomers, since the free energy is
extensive. 

The same procedure can be employed for hard gels. The relevant free energy
takes a similar form as before, apart from the elastic part of the
energy.
\begin{equation}
F\simeq\frac{M^{2}R_{\parallel}^{2}}{a^{2}N}+a^{3}\frac{N^{2}}
{HR_{\parallel}^{2}}
\end{equation}
 The size of the gel parallel to the plates is given for
completeness. It scales as 
$R_{\parallel} \simeq H (a/H)^{3/4}(N^3/M^2)^{1/4}$. To
find the confinement free energy the
same argument yields
\begin{equation}
F_{\rm conf}\simeq\left( \frac{a}{H}\right) ^{5/3}\frac{N}{M^{2/3}}
\end{equation}
An important observation is 
that for $M=1$ (no crosslink) linear chain is recovered.

\section{Adsorption behavior}
Finally we are in the position to discuss the adsorption behavior of the
gels. To begin with, we employ the same scaling arguments as given in the case
of self-similar polymeric fractal. Thus we have to consider the
competition between 
confinement, or the entropy penalty of confinement  and the 
energy gain by adsorption. This results in a total free energy of the general
form
\begin{equation}
F\simeq F_{\rm conf}-n_{a}w
\end{equation}
We just summarize the results to be brief. First for soft gels we find 
\begin{equation}
H\simeq a {w} ^{-2/3}\left( \frac{\epsilon }{a}\right)
^{2/3}M^{-1/3}
\end{equation}
Similarly for hard gels
\begin{equation}
H\simeq a {w}^{-3/2}{M}^{-1}
\end{equation}
Note that the latter equation contains the free chain result for $M=1$, 
i.e., if
no crosslinks are present.
We see that in both cases the height of the adsorbed layer depends on the
number of crosslinks in a significant and characteristic way. 
The results are in
accordance with the physical
intuition. The soft microgels adsorb more easily, because
these objects are more flexible. This is also shown by the different exponents
of the interaction energy $w$.

\section{Summary and Conclusion}

With use of Flory-Approximation we had investigated behavior of microgels. The
complexity of the distribution function of the non interacting network
prevented us to use more refined methods, as they are well known in the case of
linear polymer chains. Nevertheless we got results which are 
reasonable and could
be checked by experimental methods, at least in their tendency. The cases
worked out here have been relevant for penetrable surfaces, i.e.,
interfaces. A direct comparison to hard surfaces is not possible, since the
number of monomers close to the surface cannot be determined 
by eq. (\ref{mono}). The case of linear chains in half space has been studied
in detail, see e.g. \cite{eisenriegler} for a general reference. Crossover
exponents and new critical points determine the physics. In the present case
of microgels and polymeric manifolds a similar treatment appears very
difficult, since the "bare propagator" has a very complicated structure,
although it is exactly known \cite{solf1,solf2}. Nevertheless we expect that
the principal statements can be compared at least qualitatively with
experiments. 

To study the adsorption behavior we first
had to calculate the size of the microgel in solution. This has been carried
out by employing the Flory arguments. The basis for the reliability of the
result has been their agreement withe comparison of 
exact calculations without
excluded volume in bulk. Then 
generalization opened  the determination of sizes of microgels
with excluded volume interaction in bulk systems and
near adsorbing flat surfaces.
Moreover we made straightforward generalizations to fractal 
type microgels, which
could be described by the spectral dimension. 

The results show a transition from polymeric type of adsorption behavior for
soft microgels, i.e., with small number of crosslinks, or alternatively low
spectral dimension, to colloidal adsorption, whenever the crosslink number is
large and their coupling is strong. 

We have seen that the results have been presented from 
considerations on single gels. This has also experimental  interest on surface
protection. A layer of adsorbed gels of height $H$ at a surface is a single
gel problem. Unlike linear polymer chains the 
gels can no longer interpenetrate 
each other. Thus microgels and branched polymers
appear more effective in surface protection.  

Another interesting question is also the interplay between vulcanization and
adsorption. In the case we studied so far, we had assumed preformed gels and
fractals which had then brought to the interacting surfaces. The other case,
i.e., the vulcanization in presence of interacting walls would lead to new
types of gels with new structures which depend on the strength of the surface
- monomer interaction \cite{cohen,lip}. 

\section*{acknowledgment}
L'un des auteurs (TAV) remercie Pierre - Gilles de Gennes et Elie Rapha\"el
pour leurs remarques interessantes sur ce travail. Nous remercions aussi l'un
des critiques pour avoir attir\'e notre attention sur une methode \'el\'egante
d'obtantion du "blob", qui a permis d'ameliorer la qualit\'e de ce papier.


%
%



%


\end{document}